\documentclass[aps,twocolumn,showpacs,prb,superscriptaddress]{revtex4}

\usepackage{dcolumn}
\usepackage{amsmath}
\usepackage{epsf}
\usepackage{graphicx}

\begin{document}

\title{Planar spin-transfer device with a dynamic polarizer.}

\author{Ya.\ B. Bazaliy}
\affiliation{IBM Almaden Research Center, 650 Harry Road, San Jose, CA 95120}
\author{D. Olaosebikan}
\affiliation{Department of Physics, Cornell University, Ithaca, NY
14853}
\author{B. A Jones}
\affiliation{IBM Almaden Research Center, 650 Harry Road, San Jose, CA 95120}

\date{July, 2006}

\begin{abstract}
In planar nano-magnetic devices magnetization direction is kept
close to a given plane by the large easy-plane magnetic anisotropy,
for example by the shape anisotropy in a thin film. In this case
magnetization shows effectively in-plane dynamics with only one
angle required for its description. Moreover, the motion can become
overdamped even for small values of Gilbert damping. We derive the
equations of effective in-plane dynamics in the presence of
spin-transfer torques. The simplifications achieved in the
overdamped regime allow to study systems with several dynamic
magnetic pieces (``free layers''). A transition from a spin-transfer
device with a static polarizer to a device with two equivalent
magnets is observed. When the size difference between the magnets is
less than critical, the device does not exhibit switching, but goes
directly into the ``windmill'' precession state.
\end{abstract}


\maketitle

\section{Introduction}

The prediction \cite{berger,slon96} and first experimental
observations
\cite{berger-exp-dw,tsoi-prl-98,sun-JMMM-99,wegrowe-europhys-lett-99,ralph-science-99,ralph-prl}
of spin-transfer torques opened a new field in magnetism which
studies non-equilibrium magnetic interactions induced by electric
current. Since such interactions are relatively significant only in
very small structures, the topic is a part of nano-magnetism. The
current-induced switching of magnetic devices achieved through
spin-transfer torques is a candidate for being used as a writing
process in magnetic random access memory (MRAM) devices. The MRAM
memory cell is a typical example of a spintronic device in which the
electron spin is used to achieve useful logic, memory or other
operations normally performed by electronic circuits.

To produce the spin-transfer torques, electric currents have to flow
through the spatially non-uniform magnetic configurations in which
the variation of magnetization can be either continuous or abrupt.
The first case is usually experimentally realized in magnetic domain
walls.\cite{berger-exp-dw,saitoh-2004,klaui-2005,masamitsu-2006}
Here we will be focusing on the second case realized in the
artificially grown nano-structures. Such spin-transfer devices
contain several magnetic pieces separated by non-magnetic metal
spacers allowing for arbitrary angles between the magnetic moments
of the pieces. Magnetization my vary within each piece as well, but
that variation is usually much smaller and vanishes as the size of
piece is reduced, or for larger values of spin-stiffness of magnetic
material. The typical examples of a system with discrete variation
of magnetization are the ``nano-pillar'' devices \cite{ralph-prl}
(Fig.1A). Their behavior can be reasonably well approximated by
assuming that magnetic pieces are mono-domain, each described by a
single magnetization vector ${\vec M}(t) = M_s {\vec n}(t)$ where
$\vec n$ is the unit vector and $M_s$ is the saturation
magnetization. The evolution of ${\vec n}(t)$ is governed by the
Landau-Lifshitz-Gilbert (LLG) equation with spin-transfer
terms.\cite{slon96,bjz2000}

\begin{figure}[b]
    \resizebox{.45\textwidth}{!}{\includegraphics{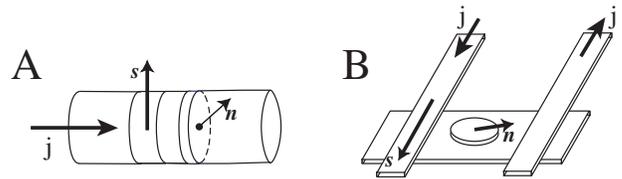}}
\caption{Planar spin-transfer devices}
 \label{fig:devices}
\end{figure}

It is often the case that magnetic pieces in a spin-transfer device
have a strong easy-plane anisotropy. For example, in nano-pillars
both the polarizer and the free magnetic layer are disks with the
diameter much larger than the thickness. Consequently, the shape
anisotropy makes the plane of the disk an easy magnetic plane. In
the planar devices \cite{bauer-planar-review} built from thin film
layers (Fig. 1B)  the shape anisotropy produces the same effect.
When the easy-plane anisotropy energy is much larger then all other
energies, the deviations of ${\vec n}(t)$ from the in-plane
direction are very small. An approximation based on such smallness
is possible and provides an effective description of the magnetic
dynamics in terms of the direction of the projection of ${\vec
n}(t)$ on the easy plane, i.e. in terms of one azimuthal angle. In
this paper we derive the equations for effective in-plane motion in
the presence of spin-transfer effect and discuss their use by
considering several examples.

In the absence of spin-transfer effects the large easy-plane
anisotropy creates a regime of overdamped motion even for the small
values of Gilbert damping constant $\alpha \ll 1$.\cite{weinan-e} In
that regime the equations simplify further. Here the overdamped
regime is discussed in the presence of electric current. The
reduction of the number of equations allows for a simple
consideration of a spin-transfer device with two dynamic magnetic
pieces. We show how an asymmetry in the sizes of these pieces
creates a transition between the polarizer-analyzer (``fixed layer -
free layer'') operation regime
\cite{slon96,ralph-prl,bjz2000,ralph-nature-2003} and the regime of
nearly identical pieces where current leads not to switching, but
directly to the Slonczewski ``windmill'' dynamic state.\cite{slon96}
Finally, we point out the limitations of the overdamped
approximation in the presence of the spin-transfer torques.

\section{Dynamic equations in the limit of a large easy-plane anisotropy}

Magnetization dynamics in the presence of electric current is
governed by the LLG equation with the spin-transfer term.
\cite{slon96,bjz2000} For each of the magnets in the device shown on
Fig.~1A
\begin{equation}
 \label{eq:vector_LLG}
{\dot {\vec n}} = \frac{\gamma}{M_s} \left[ - \frac{\delta E}{\delta
{\vec n}} \times {\vec n} \right] + u [{\vec n} \times [{\vec s}
\times {\vec n}]] + \alpha [{\vec n} \times \dot {\vec n}]
\end{equation}
where ${\vec s}\,(t)$ is the unit vector along the instantaneous
magnetization of the other magnet and the spin-transfer magnitude
\begin{equation}
 \label{eq:u_expression}
u = g(P)\frac{\gamma (\hbar/2)}{V M_s} \frac{I}{e}
\end{equation}
is proportional to the electric current $I$. Here $e$ is the
(negative) electron charge, so $u$ is positive when electrons flow
into the magnet. Due to the inverse proportionality to the volume
$V$, the larger magnets become less sensitive to the current and can
serve as spin-polarizers with a fixed magnetization direction. As
for the other parameters, $\gamma$ is the gyromagnetic ratio, $g(P,
({\vec n} \cdot {\vec s}))$ is the Slonczewski spin polarization
factor \cite{slon96} which depends on many system parameters,
\cite{brouwer,bauer_rmp} and $\alpha$ is the Gilbert damping which
also depends on ${\vec n}$ and ${\vec s}$ when spin pumping
\cite{tserkovnyak_addamping} is taken into account. We will restrict
our treatment to the constant $g$ and $\alpha$ to focus on the
effects specific to the strong easy plane anisotropy.

In terms of the polar angles $(\theta,\phi)$ the LLG equation
(\ref{eq:vector_LLG}) has the form
\begin{eqnarray}
 \nonumber
\dot\theta + \alpha \dot\phi \sin\theta &=&
    - \frac{\gamma}{M\sin\theta} \frac{\partial E}{\partial\phi}
    + u ({\vec s} \cdot {{\vec e}_{\theta}})
 \\
 \label{eq:polar_angles_LLG}
\dot\phi \sin\theta - \alpha \dot\theta &=&
    \frac{\gamma}{M} \frac{\partial E}{\partial\theta}
    +  u ({\vec s} \cdot {\vec e}_{\phi})
\end{eqnarray}
where the tangent unit vectors ${\vec e}_{\theta}$ and ${\vec
e}_{\phi}$ are defined in
Appendix~\ref{appendix:vector_definitions}.

We will consider a model for which the energy of a magnet is given
by
\begin{equation}
 \label{eq:energy}
E = \frac{K_{\perp}\cos^2\theta}{2} + E_{r}(\phi)
\end{equation}
with $K_{\perp}$ being the easy-plane constant, $E_r$ being the
``residual'' in-plane anisotropy energy and $z$-axis directed
perpendicular to the easy plane. The limit of a strong easy-plane
anisotropy is achieved when the maximal variation of the residual
energy is small compared to the easy-plane energy, $\Delta E_r \ll
K_{\perp}$. In this case $\theta = \pi/2 + \delta\theta$ with
$\delta\theta \ll 1$.

To estimate $\delta\theta$, consider the motion of magnetization
initially lying in-plane off the minimum of $E_r$ and neglect for
the moment the spin-transfer terms in
Eq.(\ref{eq:polar_angles_LLG}). Magnetization starts moving and a
certain deviation from the easy plane is developed. For the
estimate, assume that the energy is conserved during this motion
(the presence of damping will only decrease $\delta\theta$). Then
\begin{equation}
 |\delta\theta| \sim \sqrt{\frac{\Delta E_r}{K_{\perp}}} \ll 1
\end{equation}
We can now linearize the right hand sides of equations
(\ref{eq:polar_angles_LLG}) in small $\delta\theta$. On top of that,
some terms on the left hand sides of (\ref{eq:polar_angles_LLG})
turn out to be small and can be discarded. Indeed, taking into
account the smallness of $\alpha$ one gets the estimates
\begin{eqnarray}
 \nonumber
\dot\theta & \sim & -\frac{\gamma}{M_s} \frac{\partial E_r}{\partial
\phi} \sim -\frac{\gamma}{M_s} \Delta E_r
 \\
 \nonumber
\dot\phi & \sim & \frac{\gamma}{M_s} K_{\perp} \delta\theta \sim
\frac{\gamma}{M_s} \sqrt{K_{\perp} \Delta E_r}
\end{eqnarray}
Consequently $\dot\theta \sim \dot\phi\sqrt{\Delta E_r/K_{\perp}}
\ll \dot\phi$ and $\dot\phi \gg \alpha\dot\theta$, therefore the
second term on the left hand side of the second equation of the
system (\ref{eq:polar_angles_LLG}) can be discarded. No
simplification happens on the left hand side of the first equation,
where $\dot\theta$ and $\alpha\dot\phi$ can be of the same order
when $\alpha \lesssim \sqrt{\Delta E_r/K_{\perp}}$.

Putting the spin-transfer terms back we get the form of equations in
the limit of large easy-plane anisotropy:
\begin{eqnarray}
 \nonumber
\dot\delta\theta + \alpha \dot\phi &=&
    - \frac{\gamma}{M_s} \frac{\partial E}{\partial\phi}
    + u ({\vec s} \cdot {{\vec e}_{\theta}})
 \\
 \label{eq:polar_angles_LLG_approximate}
\dot\phi &=&
    \frac{\gamma K_{\perp}}{M_s} \delta\theta
    +  u ({\vec s} \cdot {\vec e}_{\phi})
\end{eqnarray}
Expressions for the scalar products in
(\ref{eq:polar_angles_LLG_approximate}) in terms of polar angles are
given in Appendix~\ref{appendix:vector_definitions}.

The second equation shows that $\delta\theta$ can be expressed
through $(\phi, \dot\phi)$. Small out-of-plane deviation becomes a
``slave'' of the in-plane motion.\cite{weinan-e} We get
\begin{equation}
 \label{eq:underdamped}
 \frac{M_s}{\gamma K_{\perp}}
 \left(
 \ddot\phi -
 u \frac{d ({\vec s} \cdot {\vec e}_{\phi})}{d t}
 \right) + \alpha_i \dot\phi  =
 - \frac{\gamma}{M_s} \frac{\partial E_{r}}{\partial \phi}
 + u ({\vec s} \cdot {{\vec e}_{\theta}})
\end{equation}
The term with the second time derivative $\ddot\phi$ decreases with
increasing $K_{\perp}$. As pointed out in
Ref.~\onlinecite{weinan-e}, in the absence of spin-transfer this
term can be neglected when $K_{\perp} > \Delta E_r/\alpha^2$.
Mathematically this corresponds to a transition from an underdamped
to an overdamped behavior of an oscillator  as the oscillator mass
decreases.

With spin-transfer terms the overdamped approximation gives an
equation
\begin{eqnarray}
 \label{eq:overdamped_1}
&& \alpha \dot\phi - \xi
  \frac{d}{d t} ({\vec s} \cdot {\vec e}_{\phi})
 = - \frac{\gamma}{M_s} \frac{\partial E_{r}}{\partial \phi}
 + u ({\vec s} \cdot {{\vec e}_{\theta}})
\end{eqnarray}
where $\xi = u M_s/(\gamma K_{\perp})$. The range of this equation's
validity will be discussed in Sec.~\ref{sec:remarks}. The scalar
products in Eq.~(\ref{eq:overdamped_1}) have to be expressed through
the polar angles $(\theta_s(t),\phi_s(t))$ of vector ${\vec s}$, and
linearized with respect to $\delta\theta$ (see Appendix, Eq.
\ref{eq:scalar_products_linear_theta_i}), which is then substituted
from Eq.~(\ref{eq:polar_angles_LLG_approximate}). Finally, the
equation is linearized with respect to small spin-transfer magnitude
$u$. We get:
\begin{eqnarray}
 \nonumber
&& \alpha \dot\phi - \xi \left(
  \frac{d}{d t} \big[\sin\theta_s\sin(\phi_s - \phi)\big]
  - \sin\theta_s\cos(\phi_s - \phi) \dot\phi
  \right)
 \\
 \label{eq:overdamped_2}
&& = - \frac{\gamma}{M_s} \frac{\partial E_{r}}{\partial \phi} -
u\cos\theta_s \ ,
\end{eqnarray}
describing the in-plane overdamped motion of an analyzer with a
polarizer pointed in the arbitrary direction. Next, we show how some
known results on spin-transfer systems are recovered in the
approximation (\ref{eq:overdamped_2}).

Consider the device shown on Fig.~1A and assume that the first
magnet is very large. As explained above, this magnet is not
affected by the current and serves as a fixed source of
spin-polarized electrons for the second magnet called the analyzer,
or the ``free'' layer. The magnetization dynamics of the analyzer is
described by Eq.~(\ref{eq:polar_angles_LLG}). The case of static
polarizer is extensively studied in the literature.

\begin{figure}[t]
    \resizebox{.45\textwidth}{!}{\includegraphics{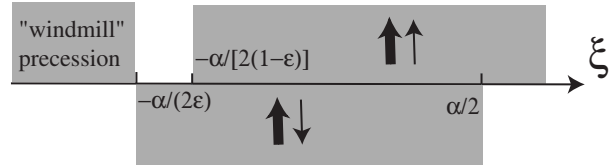}}
\caption{Stability regions for systems with static (A) and dynamic
(B) polarizers as a function of applied current, $\xi = g(P)
(\hbar/2VK_{\perp}) I/e \propto I$.}
 \label{fig:fixed_polarizer_stability}
\end{figure}

First, consider the case of {\em collinear switching},
experimentally realized in a nano-pillar device with the analyzer's
and polarizer's easy axes along the $\hat x$ direction: $E_{r} =
(1/2)K_{||} \sin^2\phi$, ${\vec s} =
(1,0,0)$.\cite{ralph-science-99} Using Eq.~(\ref{eq:overdamped_2})
with $\theta_s = \pi/2$, $\phi_s = 0$ we get
\begin{equation}
 \label{eq:collinear_equation}
(\alpha + 2 \xi\cos\phi)\dot\phi = - \frac{\gamma K_{||}}{2M_s} \sin
2\phi
\end{equation}
Without the current, there are four possible equilibria of the
analyzer. Two stable equilibria are the parallel ($\phi = 0$) and
anti-parallel ($\phi = \pi$) states. Two perpendicular equilibria
$(\phi = \pm \pi/2)$ are unstable. Linearizing
Eq.~(\ref{eq:collinear_equation}) near equilibria one finds
solutions the form $\delta\phi(t) \sim \exp(\omega t)$ with
eigenfrequencies
\begin{eqnarray*}
\omega &=& - \frac{\gamma K_{||}}{M_s (\alpha + 2 \xi)} , \qquad
(\phi \approx 0)
 \\
\omega &=& - \frac{\gamma K_{||}}{M_s (\alpha - 2 \xi)}, \qquad
(\phi \approx \pi)
 \\
\omega &=& \frac{\gamma K_{||}}{M_s \alpha},
 \qquad (\phi \approx \pm \pi/2)
\end{eqnarray*}
The equilibria are stable for $\omega < 0$ and unstable otherwise.
Thus the parallel state is stable for $\xi >  -\alpha/2$, the
antiparallel state is stable for $\xi < \alpha/2$, and the
perpendicular states cannot be stabilized by the current. These
conclusions agree with the results of
Refs.~\onlinecite{slon96,ralph-science-99,bjz2000}. The stability
regions are shown in Fig.~\ref{fig:fixed_polarizer_stability}A.

Note how Eq.~(\ref{eq:collinear_equation}) emphasizes the fact that
spin-transfer torque destabilizes the equilibria by making the
effective damping constant $\alpha_{eff} = \alpha + 2\xi\cos\phi$
negative, while the equilibrium points remain a minimum of the
energy $E_r$. Any appreciable influence of the current on the
position and nature (minimum or maximum) of the equilibrium can only
be observed at the current magnitudes $1/\alpha$ times larger than
the actual switching current.\cite{bjz2000}

Second, consider the case of {\em magnetic
fan}.\cite{axelhoffmann-magn-fan} Here the easy axis of the
polarizer is again directed along $\hat x$, but the polarizer is
perpendicular to the easy plane: ${\vec s} = (0,0,1)$, $\theta_s =
0$. This arrangement is known to produce a constant precession of
vector ${\vec n}$. Eq.~(\ref{eq:overdamped_2}) gets a form:
\begin{equation}  \label{eq:magnetic_fan}
\alpha \dot\phi = - \frac{\gamma K_{||}}{2M_s} \sin2\phi - u
\end{equation}
for $|u| < \gamma K_{||}/(2M_s)$ the current deflects the analyzer
direction from the easy axis direction. For larger values of $u$
there is no time-independent solution. The angles $\phi$ grows with
time which corresponds to ${\vec n}$ making full rotations. At $|u|
\gg \gamma K_{||}/(2M_s)$ the rotation frequency of the magnetic fan
is given by $\omega \sim u/\alpha$.

\section{Device with two dynamic magnets (two ``free layers'')}

No let us assume that both magnets in Fig.~1A have finite size. Each
magnet serves as a polarizer for the other one. Without
approximations, the evolution of two sets of polar angles
$(\theta_i,\phi_i)$, $i = 1,2$ is described by two LLG systems of
equations
\begin{eqnarray}
 \nonumber
\dot\theta^{(i)} + \alpha_i \dot\phi^{(i)} \sin\theta^{(i)} &=&
    - \frac{\gamma}{M_{si}\sin\theta^{(i)}} \frac{\partial E^{(i)}}{\partial\phi^{(i)}}
    +
 \\
 \label{eq:2_sets_polar_angles_LLG}
&+& u_{ji} ({\vec n}^{(j)} \cdot {{\vec e}_{\theta}}^{\, (i)})
 \\
 \nonumber
\dot\phi^{(i)} \sin\theta^{(i)} - \alpha_i \dot\theta^{(i)} &=&
    \frac{\gamma}{M_{si}} \frac{\partial E^{(i)}}{\partial\theta^{(i)}}
    +  u_{ji} ({\vec n}^{(j)} \cdot {\vec e}_{\phi}^{\, (i)})
\end{eqnarray}
where $j$ means the index not equal to $i$ and no summation is
implied.

We now apply the overdamped, large easy-plane anisotropy
approximation to both magnets. Equation (\ref{eq:overdamped_2}) for
each magnet is further simplified since for the magnet $i$ the angle
$\theta_s = \theta_j = \pi/2 + \delta\theta_j$, $\delta\theta_j \ll
1$. Expanding (\ref{eq:overdamped_2}) in small $\delta\theta_j$ and
using the slave condition (\ref{eq:polar_angles_LLG_approximate})
for $\delta\theta_j$ with $({\vec s} \cdot {\vec e}_{\phi}) = ({\vec
n^{(j)}} \cdot {\vec e}^{\,(i)}_{\phi})$ expanded in both small
angles (see Eq.~(\ref{eq:scalar_products_linear_theta_i_and j})) we
get the system:
\begin{eqnarray}
 \label{eq:overdamped_LLG_two_magnets}
 && (\alpha_i + 2 \xi_{ji} \cos(\phi_j - \phi_i))\dot\phi_i -
 \\
 \nonumber
 && \qquad -\xi_{ji}
(\cos(\phi_j - \phi_i) + 1)\dot\phi_j = -\frac{\partial
E^{(i)}}{\partial \phi_i} \ ,
\end{eqnarray}
with $\xi_{ji} =  u_{ji} M_{si}/(\gamma K_{\perp})$. It was assumed
that $K_{\perp}$ is the same for both magnets.

The spin-transfer torque parameters $u_{21}$ and $u_{12}$ have
opposite signs and their absolute values are different due to
different volumes of the magnets, according to
Eq.~(\ref{eq:u_expression}). We assume $V_1 \geq V_2$ and denote
$u_{12} = u$, $u_{21} = - \epsilon u$. The larger magnet experiences
a relatively smaller spin transfer effect, and the asymmetry
parameter satisfies $0 \leq \epsilon \leq 1$. In general, material
parameters $\alpha_{1,2}$, $M_{s1,2}$ and magnetic anisotropy
energies $E^{(1,2)}$ of the two magnets are also different, but here
we focus solely on the asymmetry in spin-transfer parameters. Both
$E^{(1)}$ and $E^{(2)}$ are assumed to be given by formula
(\ref{eq:energy}) with the same direction of in-plane easy axis. The
situation can be viewed as a collinear switching setup with dynamic
polarizer. Equations (\ref{eq:overdamped_LLG_two_magnets})
specialize to
\begin{eqnarray}
 \nonumber
 && \left|\begin{array}{cc}
 \alpha - 2 \epsilon\xi C &
 \epsilon \xi (C+1)
 \\
 -\xi (C+1) &
 \alpha + 2\xi C
\end{array}\right|
 \left[ \begin{array}{c}
 \dot\phi_1 \\ \dot\phi_2
 \end{array}\right] =
 - \frac{\omega_0}{2}
 \left[ \begin{array}{c}
 \sin 2\phi_1 \\ \sin 2\phi_2
 \end{array}\right]
 \\
 \label{eq:overdamped_LLG_two_magnets_easy_axis}
 && C = \cos(\phi_1 - \phi_2),
 \quad \omega_0 = \frac{\gamma
 K_{||}}{M_s}
\end{eqnarray}

Next, we study the stability of all equilibrium configurations
$(\phi_1, \phi_2)$ of two magnets. There are four equilibrium states
that are stable without the current: two parallel states along the
easy axis $(0,0)$ and $(\pi, \pi)$, two antiparallel states along
the easy axis $(0, \pi)$ and $(\pi, 0)$. Four more equilibrium
states have magnetization perpendicular to the easy axis and are
unstable without the current: $(\pm\pi/2, \pm\pi/2)$. Once again,
since spin-transfer does not depend on the relative direction of
current and magnetization, the configurations which can be
transformed into each other by a rotation of the magnetic space as a
whole behave identically. Thus it is enough to consider four
configurations: $(0,0)$, $(0, \pi)$, $(\pi/2, \pi/2)$, and $(\pi/2,
-\pi/2)$. We linearize equations
(\ref{eq:overdamped_LLG_two_magnets_easy_axis}) near each
equilibrium and search for the solution in the form $\delta\phi_i
\sim \exp(\omega t)$. The eigenfrequencies are found to be:
\begin{eqnarray}
 \nonumber
 (0,0): & & \omega_1 = \frac{-\omega_0}{\alpha},
    \quad \omega_2 = -\frac{-\omega_0}{\alpha + 2\xi(1-\epsilon)}
 \\
 \nonumber
 (0,\pi): & & \omega_1 = \frac{-\omega_0}{\alpha + 2\epsilon\xi},
    \quad \omega_2 = \frac{-\omega_0}{\alpha - 2\xi}
 \\
 \nonumber
 (\frac{\pi}{2},\frac{\pi}{2}): & & \omega_1 = \frac{\omega_0}{\alpha},
    \quad \omega_2 = \frac{\omega_0}{\alpha - 2\xi(1-\epsilon)}
 \\
 \nonumber
 (\frac{\pi}{2},-\frac{\pi}{2}): & & \omega_1 = \frac{\omega_0}{\alpha + 2\epsilon\xi},
    \quad \omega_2 = \frac{\omega_0}{\alpha - 2\xi}
\end{eqnarray}
The state is stable when both eigenfrequencies are negative. We
conclude that initially unstable states $(\pi/2, \pm\pi/2)$ are
never stabilized by the current, while the $(0,0)$ and $(0,\pi)$
state remain stable for
\begin{eqnarray}
 \nonumber
(0,0): & & \xi > -\frac{\alpha}{2(1-\epsilon)}
 \\
 \nonumber
(0,\pi): & &  -\frac{\alpha}{2\epsilon} < \xi < \frac{\alpha}{2}
\end{eqnarray}
These regions of stability are shown schematically in
Fig.~\ref{fig:fixed_polarizer_stability}B in comparison with the
case of static magnetic polarizer
(Fig.~\ref{fig:fixed_polarizer_stability}A) which is recovered at
$\epsilon \to 0$.

As the size of the polarizer is reduced, the asymmetry parameter
$\epsilon$ grows. The stability region of the antiparallel state
acquires a lower boundary $\xi = -\alpha/(2 \epsilon)$. Up to
$\epsilon = 1/2$, this boundary is still below the lower
boundary of the parallel configuration stability region.
Consequently, the parallel configuration is switched to the
antiparallel at a negative current $\xi = -\alpha/(2(1-\epsilon))$.
The system then remains in the antiparallel state down to $\xi =
-\alpha/(2\epsilon)$. Below that threshold no stable configurations
exist, and the system goes into some type of precession state. This
dynamic state is related to the ``windmill'' state predicted in
Ref.~\onlinecite{slon96} for two identical magnets in the absence of
anisotropies. Obviously, here it is modified by the strong
easy-plane anisotropy.

The $\epsilon = 1/2$ value represents a transition point in the
behavior of the system. For $1/2 < \epsilon < 1$, the stability
region of the parallel configuration completely covers the one of
the antiparallel state. A transition without hysteresis now happens
at $\xi = -\alpha/(2(1-\epsilon))$ between the parallel state and
the precession state. If the system is initially in the antiparallel
state, it switches to the parallel state either at a negative
current $\xi = -\alpha/(2(1-\epsilon))$ or at a positive current
$\xi = \alpha/2$, and never returns to the antiparallel state after
that.

\section{Concluding remarks}\label{sec:remarks}

We studied the behavior of planar spin-transfer devices with
magnetic energy dominated by the large easy-plane anisotropy. The
overdamped approximation in the presence of current-induced torque
was derived and checked against the cases already discussed in the
literature. In the new ``dynamic polarizer'' case, we found a
transition between two regimes with different switching sequences.
The large asymmetry regime is similar to the case of static
polarizer and shows hysteretic switching between the parallel and
antiparallel configurations, while in the small asymmetry regime the
magnets do not switch, but go directly into the ``windmill''
precession state.

%

We saw that the current-induced switching occurs when the effective
damping constant vanishes near a particular equilibrium. This makes
the overdamped approximation inapplicable in the immediate vicinity
of the transition and renders
Eqs.~(\ref{eq:overdamped_LLG_two_magnets_easy_axis}) ill-defined at
some points. However, the overall conclusions about the switching
events will remain the same as long as the interval of
inapplicability is small enough.

We also find that the overdamped planar approximation does not work
well when a saddle point of magnetic energy is stabilized by
spin-transfer torque, e.g. during the operation of a spin-flip
transistor. \cite{bauer-spin-flip-transistor}  Description of such
cases in terms of effective planar equations requires additional
investigations.

\section{Acknowledgements}
We wish to thank Tom Silva, Oleg Tchernyshyov, Oleg Tretiakov, and
G. E. W. Bauer for illuminating discussions. This work was supported
in part by DMEA contract No. H94003-04-2-0404, Ya.~B. is grateful to
KITP Santa Barbara for hospitality and support under NSF grant No.
PHY99-07949. D.~O. was supported in part by the IBM undergraduate
student internship program.

\appendix

\section{Vector definitions}\label{appendix:vector_definitions}

\begin{figure}[h]
    \resizebox{.25\textwidth}{!}{\includegraphics{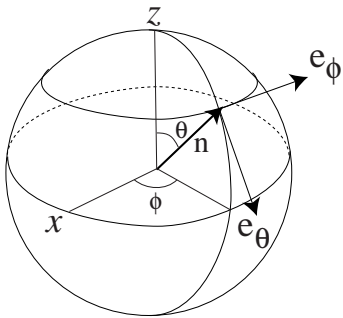}}
\caption{Definitions of the tangent vectors and polar angles.}
 \label{fig:vectors}
\end{figure}

We use the standard definitions of polar coordinates and tangent
vectors (see Fig.~\ref{fig:vectors}):
\begin{eqnarray}
 \nonumber
{\vec n} &=& (\sin\theta\cos\phi, \sin\theta\sin\phi, \cos\theta)
 \\
 \label{eq:vector_components}
{\vec e}_{\theta} &=& (\cos\theta\cos\phi, \cos\theta\sin\phi,
-\sin\theta)
 \\
 \nonumber
{\vec e}_{\phi} &=& (-\sin\phi, \cos\phi, 0)
\end{eqnarray}
When $\theta = \pi/2 + \delta\theta$ a linearization in
$\delta\theta$ gives
\begin{eqnarray}
 \nonumber
{\vec n} & \approx & (\cos\phi, \sin\phi, -\delta\theta)
 \\
 \label{eq:vector_components_linearized}
{\vec e}_{\theta} & \approx & -(\delta\theta\cos\phi,
\delta\theta\sin\phi, 1)
 \\
 \nonumber
{\vec e}_{\phi} & \approx & (-\sin\phi, \cos\phi, 0)
\end{eqnarray}
For two unit vectors ${\vec n}^{(i)}$, $i = 1,2$ with polar angles
$(\theta_i,\phi_i)$ the scalar product expressions are
\begin{eqnarray}
 \nonumber
 ({\vec n}^{(j)} \cdot {\vec e}_{\theta}^{\, (i)}) &=&
 \sin\theta_j\cos\theta_i\cos(\phi_j - \phi_i) -
 \cos\theta_j\sin\theta_i
 \\
  \label{eq:scalar_products}
 ({\vec n}^{(j)} \cdot {\vec e}_{\phi}^{\, (i)}) &=&
 \sin\theta_j \sin(\phi_j - \phi_i)
\end{eqnarray}
Linearizing (\ref{eq:scalar_products}) with respect to small
$\delta\theta_i$ for arbitrary values of $\theta_j$ one gets:
\begin{eqnarray}
 \nonumber
 ({\vec n}^{(j)} \cdot {\vec e}_{\theta}^{\, (i)}) & \approx &
 - \sin\theta_j\delta\theta_i\cos(\phi_j - \phi_i) -
 \cos\theta_j
 \\
  \label{eq:scalar_products_linear_theta_i}_
 ({\vec n}^{(j)} \cdot {\vec e}_{\phi}^{\, (i)}) & \approx &
 \sin\theta_j \sin(\phi_j - \phi_i)
\end{eqnarray}
Linearization of (\ref{eq:scalar_products}) with respect to both
$\delta\theta_i$ and $\delta\theta_j$ gives
\begin{eqnarray}
 \nonumber
 ({\vec n}^{(j)} \cdot {\vec e}_{\theta}^{\, (i)}) & \approx &
 - \delta\theta_i\cos(\phi_j - \phi_i) + \delta\theta_j
 \\
  \label{eq:scalar_products_linear_theta_i_and j}_
 ({\vec n}^{(j)} \cdot {\vec e}_{\phi}^{\, (i)}) & \approx &
 \sin(\phi_j - \phi_i)
\end{eqnarray}

\end{document}